\documentclass[journal]{IEEEtran}

\usepackage{amsmath,amsfonts,amssymb,amsthm,bbm,graphicx,enumerate,mathtools,physics,marvosym,wasysym,soul,stmaryrd,courier,listings}
\usepackage[usenames,dvipsnames]{xcolor}
\usepackage[ruled,longend]{algorithm2e}
\usepackage{etoolbox}
\usepackage{comment}
\usepackage{dsfont}
\usepackage{hyperref}
\hypersetup{colorlinks,linkcolor=black,citecolor=red,urlcolor=teal}
\usepackage[noabbrev,capitalise]{cleveref}
\usepackage{listings}

\begin{document}

% \title{\texttt{LEGO\_HQEC}: Automating the Analysis, Construction,\\ 
% and Decoding of Holographic Quantum Codes}
% \author{J.~Fan\authorrefmark{1,2}, M.~Steinberg\authorrefmark{1,2}, A.~Jahn\authorrefmark{3}, C.~Cao\authorrefmark{4}, A.~Sarkar\authorrefmark{1,2}, and S.~Feld\authorrefmark{1,2}}
% \address[1]{QuTech, Delft University of Technology, 2628 CJ Delft, The Netherlands}
% \address[2]{Quantum and Computer Engineering Department, Delft University of Technology, 2628 CD Delft, The Netherlands}
% \address[3]{Department of Physics, Freie Universit\"at Berlin, 14195 Berlin, Germany}
% \address[4]{Department of Physics, Virginia Tech, Blacksburg, VA 24061, USA}

% \corresp{Corresponding author: Matthew Steinberg (email: matt.steinberg3@gmail.com).}
\title{\texttt{LEGO\_HQEC}: Automating the Analysis, Construction,\\ 
and Decoding of Holographic Quantum Codes}

\author{J.~Fan, M.~Steinberg, A.~Jahn, C.~Cao, A.~Sarkar, and S.~Feld%
\thanks{J. Fan, M. Steinberg, A. Sarkar, and S. Feld are with QuTech, Delft University of Technology, 2628 CJ Delft, The Netherlands, and with the Quantum and Computer Engineering Department, Delft University of Technology, 2628 CD Delft, The Netherlands.}%
\thanks{A. Jahn is with the Department of Physics, Freie Universit\"at Berlin, 14195 Berlin, Germany.}%
\thanks{C. Cao is with the Department of Physics, Virginia Tech, Blacksburg, VA 24061, USA.}%
\thanks{Corresponding author: M. Steinberg (e-mail: matt.steinberg3@gmail.com).}%
}

\maketitle

%--------------------------------------------%
\begin{abstract}
Quantum error correction (QEC) is a crucial prerequisite for future large-scale quantum computation. Finding and analyzing new QEC codes, along with efficient decoding and fault-tolerance protocols, is central to this effort. Holographic codes are a recent class of generalized concatenated codes derived from holographic bulk/boundary dualities. In addition to exploring the physics of such dualities, these codes possess useful QEC properties such as tunable encoding rates, distance scaling competitive with other well-studied code classes, and excellent recovery thresholds. To allow for a comprehensive analysis of holographic code constructions, we introduce \texttt{LEGO\_HQEC}, a software package utilizing the quantum LEGO formalism. This package allows for the construction and analysis of holographic codes on regular hyperbolic tilings, computing their stabilizer generators and logical operators for a specified number of seed codes and layers. Three decoders are included: an erasure decoder based on Gaussian elimination; an integer-optimization decoder; and a tensor-network decoder. With these tools, \texttt{LEGO\_HQEC} enables systematic studies of both previously known holographic codes and novel variants. As a demonstration, we provide new numerical results on the holographic black-hole pentagon code, establishing its threshold behavior under the erasure channel as a benchmark example.
\end{abstract}
%--------------------------------------------%


% \tableofcontents

%--------------------------------------------%
\section{Introduction}\label{section:intro}

The field of quantum computing is currently focused on understanding how best to construct a large-scale fault-tolerant quantum computer~\cite{campbell2017roadsterhal,gottesman2013fault,preskill2018quantum}. The physical levels of noise that dominate in current quantum computing hardware are simply too high to reliably run all but the simplest quantum algorithms. As a solution, \emph{quantum error correction} (QEC) techniques~\cite{lidar2013quantum,wilde2013quantum,nielsen2010quantum,gottesman1997stabilizer} were proposed in order to orchestrate strategies that protect information at the logical level and facilitate the reliable execution of quantum algorithms on quantum devices. 

Although QEC in theory posits a method by which practically useful quantum computing could be realized, in reality the constraints imposed by current state-of-the-art code constructions and so-called \emph{fault-tolerance} protocols are extremely resource-intensive~\cite{gottesmanFT}. As such, almost all areas of the QEC design pipeline are under active research. These research areas range from the theoretical to the practical, and include: code constructions~\cite{albert2022bosonic,breuckmann2021quantum,jahn_topical}; fault tolerant error correction protocols~\cite{between_shorsteane,dhruvi_groovy,chamberland2018flag,delfosse2021beyond,bombin2015single,divincenzo1996fault,divincenzo2007effective,chao2018quantum,aliferis2005quantum,knill2005quantum,shor1996fault,delfosse2020short}; fault-tolerant universal logical gate set implementations, such as \emph{magic-state distillation}~\cite{bravyi2005universal}, \emph{gauge-fixing}~\cite{bombin2015gauge}, and \emph{code switching}~\cite{codeswitching1,codeswitching2,codeswitching3}, among others~\cite{jochym2014using,chamberland2016thresholds}; decoder design~\cite{iolius2023decoding,dinur2023good,roffe2020decoding,babar2015fifteen}; architectural proposals~\cite{campbell2017roadsterhal,terhal2015quantum,yoshida2024concatenate}; among many other engineering-level challenges~\cite{fowler2012surface,breuckmann2021quantum}. 

Holographic quantum codes are a recently introduced class of quantum error correction codes~\cite{preskill_happy_code,harlow2015bulk,jahn_topical}, derived as models of bulk/boundary dualities in theoretical physics, in particular  the \emph{AdS/CFT correspondences}~\cite{maldacena1999large,witten1998anti}.
The main idea for these codes lies in the fact that \emph{bulk degrees of freedom}, which are associated with logical qubits in $d$ dimensions, are mapped to a $(d-1)$-dimensional \emph{boundary}, with the hyperbolic structure of the bulk ensuring that the physical Hilbert space is larger than its logical counterpart.
Constructed on hyperbolic manifold tessellations, they can be expressed as \emph{generalized concatenated codes} and allow for extremely high (and tunable) encoding rates as well as excellent threshold performance in comparison to other leading quantum code constructions, such as \emph{quantum low-density parity-check} (qLDPC)~\cite{breuckmann2021quantum} and \emph{topological} (TQEC) quantum codes~\cite{fowler2012surface,terhal2015quantum}. However, for all practical purposes, actually generating the \emph{conformal quasicrystal lattices}~\cite{conformal_quasicrystals_holography,jahn2022tensor_qcft,central_charge_aperiodic} responsible for defining physical-qubit representations of stabilizers and logical operators is a challenging task, exacerbated by the need to apply suitable decoding algorithms derived from tensor-network decoding techniques~\cite{joe_chubb,farrelly_parallel_decoding,farrelly_tn_codes,chubb2021statistical}. In this vein, having a software tool at hand to compute the effect of quantum noise channels is indispensable for the study of new holographic (as well as generalized concatenated) quantum codes.

In previous studies~\cite{farrelly_tn_codes,Farrelly_ltnc,qecsim}, tensor-network decoders have been implemented using the Julia language (which is available at \url{https://github.com/qecsim/TensorNetworkCodes.jl}). Although one can in principle also build holographic tensor-network codes using this package, our work provides a specific advantage for studying holographic codes due to its streamlined setup and its specialization to generating regular hyperbolic lattices with a quasiperiodic boundary in an efficient manner, as well as its ability to test codes under a wide range of possible noise channels; these include the erasure, depolarizing, and biased-noise channels (such as those explored in~\cite{fan2024overcoming}).

In this manuscript, we introduce a novel software tool named \texttt{LEGO\_HQEC} in order to expedite and lower the barrier to entry for this new subfield of quantum error correction. In particular, we implement a stabilizer and logical operator generator that returns a list of these for both \emph{maximum-rate} and \emph{asymptotically zero-rate} variants of many existing holographic codes, as well as providing valuable customization options for constructing \emph{homogeneous} or \emph{heterogeneous} constructions (i.e., holographic codes constructed from either one or several smaller \emph{seed codes}) on hyperbolic manifolds parameterized by \emph{Schl\"{a}fli symbols} $\{p,q{=}4\}$~\cite{ratcliffe1994foundations}. In addition to this tool, we provide three different decoders which have been used in recent studies regarding the threshold properties of holographic codes for practical quantum computing: an \emph{erasure decoder} built using a \emph{Gaussian elimination} (GE) method~\cite{steinberg2024far,harris_css_hqec}; a Pauli-noise \emph{integer-optimization} decoder, as from~\cite{harris_int_opt,steinberg2024far}; and a \emph{tensor-network} decoder, which was utilized in~\cite{fan2024overcoming,farrelly_parallel_decoding,farrelly_tn_codes} additionally for Pauli noise. With the aim of developing newer and better open-source software tools for studying quantum error correction, this package has been made publicly available at \url{https://github.com/QML-Group/HQEC}. In this way, the overarching design premise of \texttt{LEGO\_HQEC} involves \emph{automation} and \emph{extendability} for creating, analyzing, and decoding holographic quantum codes.

This paper is organized as follows: \cref{section:background} gives an overview of the \emph{quantum LEGO} formalism, and how it relates to stabilizer and holographic QEC codes; \cref{section:software_design} details the structure of \texttt{LEGO\_HQEC}, starting with boundary stabilizer and logical operator generation, and ending with a concise overview of the distinct decoders offered; in \cref{section:using_lego_hqec}, we show two examples using: the $\llbracket 5,1,3 \rrbracket$ pentagon (or \emph{Harlow-Preskill-Pastawski-Yoshida} (HaPPY)~\cite{preskill_happy_code}) code, as well as a holographic \emph{Steane} code~\cite{harris_css_hqec}. We further present novel numerical results on the black-hole HaPPY code in \cref{section:happy_erasure}, highlighting its threshold behavior under the erasure channel as a demonstration of the tool’s utility. The calculated runtime for generating stabilizers and logical operators for several layers of the asymptotically zero-rate HaPPY code is presented, as well as decoding for erasure and Pauli noise in the asymptotically zero-rate Steane code, in addition to the maximum-rate HaPPY code. We conclude by proposing several extensions to \texttt{LEGO\_HQEC} in \cref{section:conclusion}.

%--------------------------------------------%
\section{Background}\label{section:background}

%--------------------------------------------%
\subsection{Quantum Error Correction \& Tensor Networks}\label{section:QEC_TN_background}

Quantum error correction (QEC) codes are used to protect information at the level of physical qubits (or qudits) from deleterious environmental noise effects. Such protection is accomplished by redundantly mapping physical qubit states into subspaces of a larger, multipartite Hilbert space; this description at the level of an encoding isometry effectively defines a QEC code~\cite{nielsen2010quantum,lidar2013quantum}. 

There are many different methods by which one can construct a QEC code; however, for brevity and for the scope in this article, we will focus on \emph{quantum stabilizer codes}, a class of codes for which the \emph{codespace} $\mathsf{C}$ is defined by the simultaneous $+1$ eigenspace of \emph{Pauli stabilizer generators}; that is,
\begin{equation}
\mathsf{C} = \{\ket{\psi} \in \mathcal{H}: S\ket{\psi} = \ket{\psi}, \forall S \in \mathsf{S}\}~,
\end{equation}
wherein the set of generators $\{ S \}$ which stabilize the codespace form an \emph{Abelian subgroup} which is known as the \emph{stabilizer group} $\mathsf{S}$, and as such $\{ S \} \in \mathsf{S}$. 
In order to define the logical operators of a quantum code, we first must define several terms. The \emph{normalizer} $\mathcal{N}$ of the stabilizer group is defined as 
\begin{equation}
\mathcal{N}(\mathsf{S}) = \{ \mathsf{P} \in \mathsf{P}_{n}: \mathsf{P}\mathsf{S}\mathsf{P}^{\dagger} = \mathsf{S} \},
\end{equation}
where $\mathsf{P}$ is an element of the $n$-fold Pauli group $\mathsf{P}_{n}$. Then, we define the logical operators of a quantum code as pertaining to the group $\mathcal{N}(\mathsf{S}) / \mathsf{S}$. 

% \emph{Subsystem codes} are typically stabilizer codes in which the encoded subspace is split up and subdivided. Normally, the Hilbert space of a stabilizer code is divided as $\mathcal{H} = \mathcal{H}_{C} \oplus \mathcal{H}_{\bar{C}}$, where $\mathcal{H}_{C}$ is used to store the logical information, and $\mathcal{H}_{\bar{C}}$ is the \emph{complement}. We can additionally subdivide $\mathcal{H}_{C}$ further as $\mathcal{H}_{C} = \mathcal{H}_{L} \otimes \mathcal{H}_{J}$, where $\mathcal{H}_{L}$ is the \emph{logical subspace} and $\mathcal{H}_{J}$ is known as the \emph{junk} (or gauge) subspace~\cite{lidar2013quantum}. The gauge subspace is used to help diagnose and correct errors inside of $\mathcal{H}_{L}$; such a feat is accomplished by defining the gauge group such that the stabilizer subgroup is the \emph{center} $\mathsf{Z}(\mathsf{G})$ of the gauge group

% \begin{equation}
% \mathsf{Z}(\mathsf{G}) := \{ i\mathcal{I}, \mathsf{S}\} = \mathsf{C}(\mathsf{G}) \cap \mathsf{G}~,
% \end{equation}

% where $\mathsf{G}$ is the \emph{junk} (or gauge) group associated with $\mathcal{H}_{J}$, and $\mathsf{C}(\mathsf{G})$ is the \emph{centralizer} of $\mathsf{G}$, which is defined as $\mathsf{C}(\mathsf{G}) := \{\mathcal{O} \in \mathsf{G} | [\mathcal{O}, S] = 0, \forall S \in \mathsf{S}\}$.

\begin{figure}
\centering
\includegraphics[width=\columnwidth]{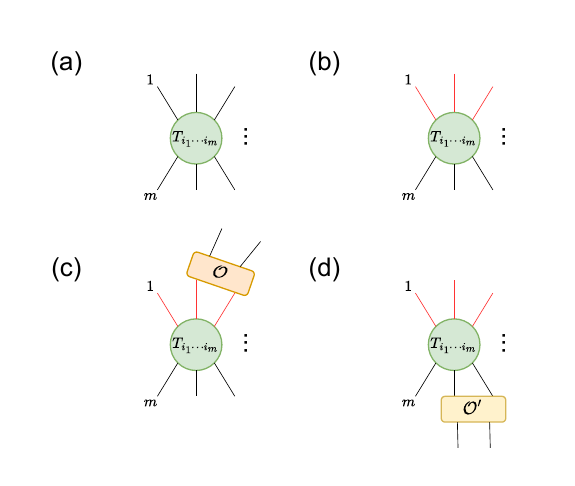}
\caption{Tensor-network diagrams depicting (a)-(b) the tensor $T_{i_{1} \cdots i_{m}}$ as it pertains to a quantum state, as well as an isometric map via the Choi-Jamiolkowski isomorphism~\cite{wilde2013quantum}. Here, red open legs represent logical input qubits, and black legs depict physical output qubits. (c)-(d) The action of tensor pushing an operator $\mathcal{O}$ through the input subsystems of the isometry in order to find an output-leg representation, which we name $\mathcal{O}'$.}
\label{fig:TN_pushing}
\end{figure}

As a specific example of a stabilizer code, we introduce here the \emph{Steane} code~\cite{steanecode1,steanecode2}, which belongs to the class of \emph{Calderbank-Shor-Steane} codes. This $\llbracket 7,1,3 \rrbracket$ code has $X$ and $Z$ stabilizers symmetrically split in the parity check matrix, and the stabilizer generators take the form
\begin{subequations}
\begin{align}
X_{1}X_{4}X_{6}X_{7}~, \\
X_{2}X_{4}X_{5}X_{7}~, \\
X_{3}X_{5}X_{6}X_{7}~, \\
Z_{1}Z_{4}Z_{6}Z_{7}~, \\
Z_{2}Z_{4}Z_{5}Z_{7}~, \\
Z_{3}Z_{5}Z_{6}Z_{7}~,
\end{align}
\end{subequations}
with logical operators defined as 
\begin{subequations}
\begin{align}
    \bar{X} &= X_{1}X_{2}X_{3}X_{4}X_{5}X_{6}X_{7} \ , \\
    \bar{Z} &= Z_{1}Z_{2}Z_{3}Z_{4}Z_{5}Z_{6}Z_{7} \ .
\end{align}
\end{subequations}
Another small example of a QEC code is the \emph{perfect} $\llbracket 5,1,3 \rrbracket$ code, which is known as the smallest code with code distance $d=3$~\cite{laflamme1996perfect}; its stabilizer generators are
\begin{subequations}
\begin{align}
X_{1}Z_{2}Z_{3}X_{4}~, \\
X_{2}Z_{3}Z_{4}X_{5}~, \\
X_{1}X_{3}Z_{4}Z_{5}~, \\
Z_{1}X_{2}X_{4}Z_{5}~,
\end{align}
\end{subequations}
with the logical operators defined as
\begin{subequations}
\begin{align}
\bar{X} &= X_{1}X_{2}X_{3}X_{4}X_{5}~, \\
\bar{Z} &= Z_{1}Z_{2}Z_{3}Z_{4}Z_{5}~.
\end{align}
\end{subequations}
These two examples in particular are important, as they constitute some of the elemental \emph{seed codes} for constructing a holographic quantum code via the \emph{quantum LEGO} (qLEGO) formalism~\cite{farrelly_tn_codes,quantumlego,Farrelly_ltnc}. More succinctly, the qLEGO formalism provides a method by which \emph{tensor networks} can be employed in order to construct and simulate large QEC codes from seed codes. One starts by defining a quantum state as 
\begin{equation}
\ket{\psi} = \sum_{i_{1} \cdots i_{m}} T_{i_{1} \cdots i_{m}} \ket{i_{1} \cdots i_{m}}~,
\end{equation}
where the tensor $T_{i_{1} \cdots i_{m}}$ also can be used to describe a mapping from input subsystems to output subsystems, as shown in \cref{fig:TN_pushing}(a)-(b). Here the repeated indices are summed over.  If qubits $1$ through $k$ are maximally entangled, we can write an \emph{isometry} for $T_{i_{1} \cdots i_{m}}$ as 
\begin{equation}
V_{\ket{\bar{\psi}}} := \sum_{i_{1} \cdots i_{m}} T_{i_{1} \cdots i_{m}}\ket{i_{k+1} \cdots i_{m}}\bra{i_{1} \cdots i_{k}}~,
\end{equation}
in which we can easily take the incoming $i_{1} \cdots i_{k}$ input legs as the encoding qudits for an $\llbracket n, k \rrbracket$ QEC code (where $n$ and $k$ represent the number of physical and logically-encoded qudits, respectively), following~\cite{quantumlego}. In general, it is not necessary that $T_{i_{1} \cdots i_{m}}$ represent an isometry~\cite{quantumlego}; however, a convenient choice is to require that $T_{i_{1} \cdots i_{m}}$ corresponds to a \emph{multipartite maximally-entangled} (MME) state~\cite{zyczkowski_mme}, thereby necessitating that a many-body quantum state adheres to the condition 
\begin{equation}
\rho_{s} = \text{Tr}_{s^{c}} \big[ \ket{\psi}\bra{\psi} \big] \propto \mathbb{I}~, \label{eq:ame}
\end{equation}
where $s \in \{1 \cdots n\}$, $|s| \leq \left \lfloor{\frac{n}{2}}\right \rfloor$, and $s^{c}$ denotes the subsystems set complementary to $s$. States adhering to this property are known as \emph{k-uniform}, and have applications in many areas of quantum information theory~\cite{zyczkowski_mme}. Note that the qLEGO formalism also applies to more general tensors that do not form isometries, as the contraction of isometric tensors (such as those above) permits the straightforward formation of a quantum code's encoding circuit~\cite{quantumlego}. 

Continuing, we find a graphical description in \cref{fig:TN_pushing}(c)-(d) of the concept known as \emph{tensor pushing}. Simply put, since $T_{i_{1} \cdots i_{m}}$ can form an isometry from $i_{1} \cdots i_{k}$ to $i_{k+1} \cdots i_{m}$, it is possible to input an operator $\mathcal{O}$ with support on the input legs (as shown) and \emph{push the operator} past the isometry, finding a new representation of the operator on the output legs. 

\begin{figure}
\centering
\includegraphics[width=1.05\columnwidth]{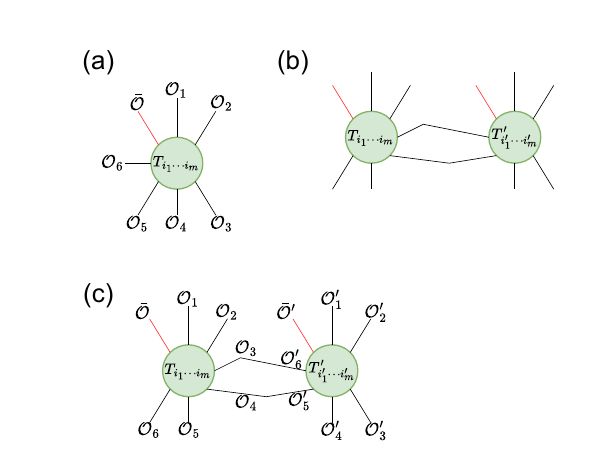}
\caption{(a) One may define an operator $\mathcal{\bar{O}}$ on an input leg to a tensor by its representation (in the present case $\mathcal{O}_{1} \otimes \cdots \otimes \mathcal{O}_{6}$) on the output legs. (b) Contracting indices together yields a new, conjoined tensor network. (c) Deducing the new logical operator is possible by keeping the original logical operators that satisfy the matching condition
$\mathcal{O}_{3} = \mathcal{O}'^{*}_{6}$ and $\mathcal{O}_{4} = \mathcal{O}'^{*}_{5}$~\cite{quantumlego,Farrelly_ltnc}.}
\label{fig:new_stabs}
\end{figure}

\cref{fig:new_stabs}(a) shows the support of a unitary logical operator $\mathcal{\bar{O}}$ on the output subsystems as $\mathcal{\bar{O}} = \mathcal{O}_{1} \otimes \cdots \otimes \mathcal{O}_{6}$ (in our specific case, for simplicity). In (b), we see that contracting two such tensors along their respective indices gives rise to a new conjoined tensor (in this case with two logical qubits). Finding the new logical operators for this new isometry is quite simple, and follows the tensor-matching formalism described in detail in~\cite{quantumlego,Farrelly_ltnc}. As shown in (c), the principle requirement is that support on contracted output indices must be identical up to complex conjugation; in the specific case of \cref{fig:new_stabs}(c), the requirements must be that $\mathcal{O}_{3} = \mathcal{O}'^{*}_{6}$ and $\mathcal{O}_{4} = \mathcal{O}'^{*}_{5}$.

The concepts and operations explained above will be of paramount use in the software implementation, as we will see in \cref{section:software_design}.

%--------------------------------------------%
\subsection{Holographic Quantum Codes}\label{section:HQEC_background}

Holographic quantum codes are a relatively new class of \emph{generalized concatenated codes}~\cite{heterogeneous_holo_qrm,steinberg2024far,steinberg_2023}. These codes have been found to exhibit properties consistent with the \emph{AdS/CFT correspondence}~\cite{harlow2015bulk,witten1998anti,maldacena1999large,gubser1998gauge}, in addition to possessing interesting and potentially useful error-correction properties for practical quantum computing, such as: high, tunable rates~\cite{steinberg2024far,farrelly_parallel_decoding}; versatility in constructions involving several different seed codes~\cite{heterogeneous_holo_qrm}; distance scaling better than $\sqrt{n}$~\cite{harris_int_opt}; and constructions with large fault-tolerant logical gate sets~\cite{heterogeneous_holo_qrm}, among other useful properties~\cite{fan2024overcoming,junyu_thesis}. 

% \clearpage
\begin{figure}
\centering
\includegraphics[width=\columnwidth]{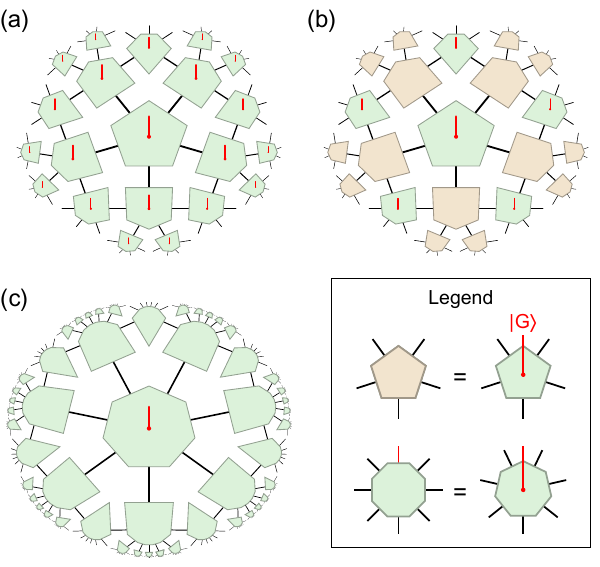}
\caption{Examples of several holographic tensor-network codes that \texttt{LEGO\_HQEC} can construct, decode, and analyze. (a) The maximum-rate \emph{Harlow-Preskill-Pastawski-Yoshida} (HaPPY) code on a hyperbolic pentagon tiling, mapping from logical (red) to physical (black) legs (\cref{section:happy_erasure}). (b) A partially gauge-fixed version of the HaPPY code is created by projecting some logical legs onto an eigenstate $\ket{G}$ of a gauge operator $G$.
(c) The asymptotically zero-rate \emph{holographic Steane code}, which uses the same 8-leg tensor once as a logical-to-physical map (central heptagon tensor) and otherwise with only planar legs (octagon tensors); a test of this code is performed in \cref{section:steane_depo}.}
\label{fig:hqecc_types}
\end{figure}

\begin{figure}
\centering
\includegraphics[width=\columnwidth]{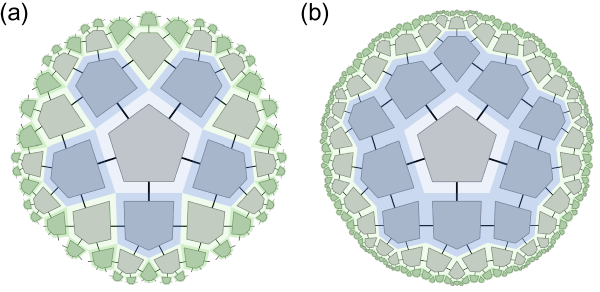}
\caption{Methods by which one can perform holographic concatenation. Both (a) edge inflation and (b) vertex inflation of a tensor network on the $\{5,4\}$ face-centered hyperbolic tessellation~\cite{conformal_quasicrystals_holography,jahn2022tensor_qcft,jahn_topical}. In both cases, we show the first three inflation steps around a central pentagon.
}
\label{fig:inflation_rules}
\end{figure}

A holographic quantum code is built typically from either one or several \emph{seed codes}, as shown in \cref{fig:hqecc_types}. As hyperbolic tessellations allow for many different orientations and polygonal embeddings in the bulk subject to the constraint $\frac{1}{p} + \frac{1}{q} < \frac{1}{2}$, infinitely many variations of holographic quantum codes are possible to construct~\cite{heterogeneous_holo_qrm,ratcliffe1994foundations}. However, only \emph{homogeneous}~\cite{harris_css_hqec,preskill_happy_code,farrelly_tn_codes,steinberg2024far} (that is, those holographic codes constructed from uniform hyperbolic tessellations with only one polygon featured, as shown in \cref{fig:hqecc_types}(a)) and simple \emph{heterogeneous} (with only alternating layers of two seed codes, as in \cref{fig:hqecc_types}(b)) holographic codes~\cite{heterogeneous_holo_qrm} have so far been constructed. Growing a holographic code layerwise is done in two distinct methods, as shown in \cref{fig:inflation_rules}: either using (a) \emph{edge inflation}, in which every inflation step adds the polygons adjacent to the edges of the previous boundary, or (b) \emph{vertex inflation}, in which each inflation step adds the polygons adjacent to the boundary vertices, resulting in a connected ring of polygons to be added.  In the tensor-network scheme, subsequent layers must contract indices of the previous layer in order to adhere to the stringent constraints of a hyperbolic tessellation. More details on this topic and the motivation behind holographic codes can be found in~\cite{jahn_topical}.

%--------------------------------------------%
\section{Software Design}\label{section:software_design}

\texttt{LEGO\_HQEC} was developed in Python. %with ease of use front and center in the design considerations. As such, \texttt{LEGO\_HQEC} 
For ease of use, the code is modularized into two parts: (a) the stabilizer and logical operator generator, and (b) the decoders. Both will be addressed in what follows.

%--------------------------------------------%
\subsection{Stabilizer \& Logical Operator Generation}\label{section:stab_logi_op_gen}

\begin{figure}
\centering
\includegraphics[width=\columnwidth]{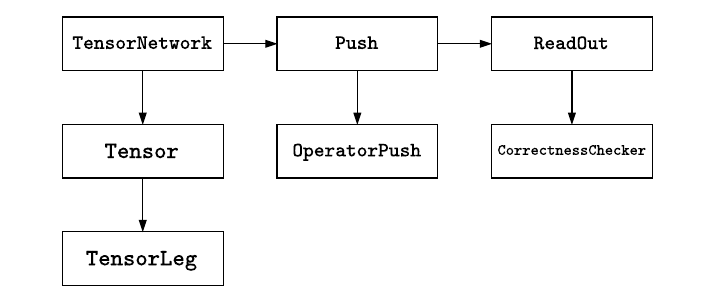}
\caption{Pictographic representation of the stabilizer / logical operator generator in \texttt{LEGO\_HQEC}. \texttt{TensorNetwork} is a list containing all of the \texttt{Tensor} objects, whose attributes include the \texttt{TensorLeg} object. This list of tensors is then filtered into the \texttt{Push} function, which elementwise processes and pushes each operator from each tensor to the boundary, obtaining a boundary representation. Finally, a correctness checker (\texttt{CorrectnessChecker}) backtracks along the tensor network, checking to see that each contracted index from each tensor contains only the identity operator. The final boundary representation is read out with \texttt{ReadOut}.}
\label{fig:stab_log_op_generator_software}
\end{figure}

Stabilizer and logical operator generation is carried out following several steps, which are depicted in \cref{fig:stab_log_op_generator_software}. Firstly, the object \texttt{TensorNetwork} is defined, with dependencies \texttt{Tensor} and \texttt{TensorLeg} as important classes; these are the components used to construct the holographic tensor network. Next, the \texttt{Push} function is defined; in this portion, each tensor has its own \texttt{OperatorPush} method on its local legs. \texttt{Push} then activates tensors one-by-one from the center to the boundary, and sequentially generates boundary representations of all locally-defined stabilizers and logical operators. Additionally, legs that have undergone an operator push will be blocked, in order to prevent operators from being accidentally pushed back in towards the center; in this way a locally-defined operator can be pushed to the boundary of the network. Finally, the \texttt{ReadOut} layer retrieves boundary stabilizers and logical operators using a backtracking method, and compiles them into the final boundary result; this backtracking method starts from the center of the tensor network, then uses a depth-first traversal to reach outer tensors, while recording operators on the legs of the outermost tensors at the boundary. With this method, it is ensured that boundary operators align with the geometric intuition expected from holographic codes. As a last step, a correctness check is performed by again traversing all internal legs and checking that all bulk-defined operators have been pushed to the boundary. More details on this portion of the software are available in~\cite{junyu_thesis}.

%--------------------------------------------%
\subsection{Decoders}\label{section:decoders}

Three decoders are included in \texttt{LEGO\_HQEC}: An erasure decoder (\cref{section:erasure_noise}) for the quantum erasure channel~\cite{wilde2013quantum,nielsen2010quantum} as well as two decoders for Pauli noise based on \emph{integer optimization} and a \emph{tensor-network} decoder (\cref{section:pauli_noise})~\cite{farrelly_parallel_decoding,farrelly_tn_codes}.

%--------------------------------------------%
\subsubsection{Quantum Erasure Channel}\label{section:erasure_noise}

\emph{Erasure} (or \emph{qubit loss}) is a quantum noise channel $\mathcal{E}_{e}(\cdot)$ with two inputs ($\ket{0}, \ket{1}$) and three outputs ($\ket{0}, \ket{1}, \ket{e}$)~\cite{wilde2013quantum,capacities_erasure,nielsen2010quantum}, and can be written as

\begin{equation}
\mathcal{E}_{e}(\rho) = (1-\epsilon)\rho + \epsilon \text{Tr}\big[ \rho \big]\ket{e}\bra{e}~,
\end{equation}

where with some probability $(1-\epsilon)$ a qubit is left unchanged and with probability $\epsilon$ the qubit is ``erased". $\ket{e}$ is a state orthogonal to the input state $\rho$.  

\texttt{LEGO\_HQEC} includes an erasure decoder based on \emph{Gaussian elimination} (GE), as was utilized in~\cite{steinberg2024far,harris_css_hqec}. The general idea of the GE decoder involves stacking all of the stabilizers and logical operators into a large matrix, and then using the support of the erasure vector, for which entries are randomly generated per qubit with probability $p$, to filter out columns that have undergone an erasure. The algorithm then proceeds by executing Gaussian elimination to find the reduced-row-echelon form of the matrix; if it is found that both logical operators still have physical support on the remaining qubits, then the trial is counted as a success. The runtime complexity of the GE algorithm is known to scale as $\mathcal{O}(n(n-k)^{2})$. More details on the GE decoder can be found at~\cite{harris_css_hqec,steinberg2024far}. 

%--------------------------------------------%
\subsubsection{Pauli Noise Channels}\label{section:pauli_noise}

The prototypical Pauli noise channel is represented by the \emph{depolarizing} noise channel $\mathcal{E}_{d}(\cdot)$~\cite{nielsen2010quantum,wilde2013quantum}, and takes the form

\begin{equation}
\mathcal{E}_{d}(\rho) = (1-\delta)\rho + \frac{\delta}{3}\big( X\rho X + Y\rho Y + Z\rho Z \big)~,
\end{equation}

where $\delta$ is the probability of an error occurring, and $\{ X, Y, Z\}$ represent the three Pauli operators. For Pauli noise, \texttt{LEGO\_HQEC} employs two different types of decoders: one based on \emph{integer optimization}~\cite{steinberg2024far,harris_int_opt}, and another based on \emph{tensor-network} decoding~\cite{fan2024overcoming,farrelly_tn_codes}. 

Integer optimization decoding works via a very similar technique to the GE decoder discussed in \cref{section:erasure_noise}, but with some small changes: Firstly, as error locations are not known, one must first calculate the \emph{Moore-Penrose pseudoinverse}~\cite{penrose1955generalized,moore1920reciprocal}, which corresponds to the \emph{destabilizers} (or pure errors) of the code~\cite{harris_int_opt,gottesman_improved_sim}. One then obtains a guessed error by multiplying the pseudoinverse matrix directly by the vector of the obtained syndrome. Finally, an integer optimization is utilized in order to deduce the lowest-weight pure-error correction~\cite{junyu_thesis,harris_int_opt}. It is known that integer-optimization decoding exhibits a runtime complexity of $\mathcal{O}(2^{n})$, which is dependent on the size of the quantum code. More details on integer-optimization decoding can be found in~\cite{junyu_thesis,steinberg2024far,harris_int_opt}.

Tensor-network decoding, however, works by exploiting the tensor-network geometry and clever contraction sequences~\cite{farrelly_tn_codes,farrelly_parallel_decoding}. By setting the bond dimension of each tensor in the network to be four, Pauli errors $\{ I, X, Y, Z\}$ can be easily accounted for. In this way, an error vector is initialized at random with some error probability $d$. These error probabilities are then utilized in the tensor network, and, upon contraction, yield a close approximation in the thermodynamic limit to \emph{maximum-likelihood decoding}~\cite{chubb2021statistical}. It is known that tensor-network decoding exhibits a runtime complexity of $\mathcal{O}(n^{2.37})$~\cite{le_gall_matrix_mult,bravyi_suchara}\footnote{Although the runtime complexity above has been reported in several studies, we note that the overall runtime complexity for tensor-network decoding depends strongly on the tensor network's geometry, as well as the contraction scheme (i.e., whether or not approximate contractions are used) and contraction ordering.}, and that multiple logical qubits can be decoded in parallel as $\mathcal{O}(nk)$, as long as the error rate stays below the critical error rate of the central logical qubit (i.e., the threshold of the central logical qubit)~\cite{farrelly_parallel_decoding}. More details on tensor-network decoding can be found in~\cite{joe_chubb,farrelly_tn_codes,farrelly_parallel_decoding}. 

%--------------------------------------------%
\section{Examples Using \texttt{LEGO\_HQEC}}\label{section:using_lego_hqec}

% \textcolor{red}{[edit this section! - Installation and Usage - give typical HaPPY code examples - updated pseudocode]}

\begin{figure*}
\centering
\includegraphics[width=0.9\textwidth]{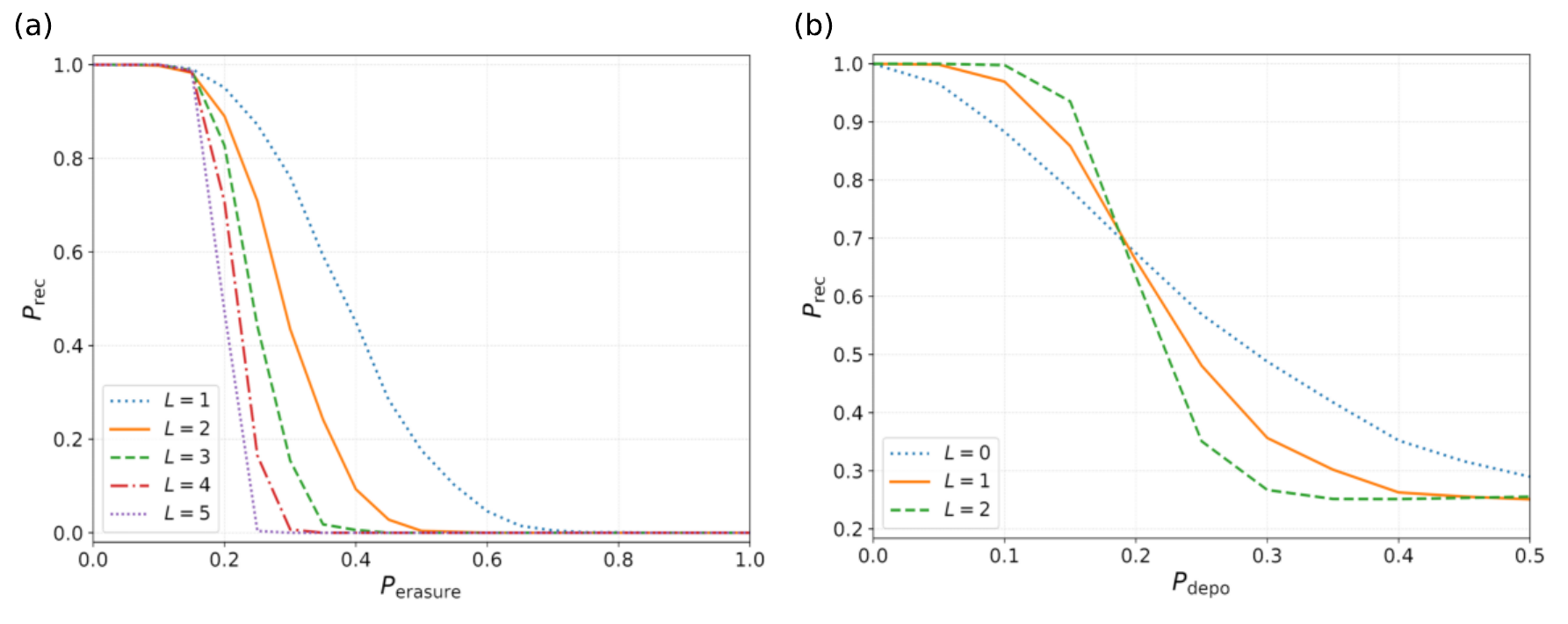}
\caption{(a) Erasure decoder results for the maximum-rate HaPPY code (which obtains no threshold, as per the results in~\cite{preskill_happy_code}), which utilized the GE decoder (\cref{section:erasure_noise}). (b) Pauli depolarizing noise results from the tensor-network decoder for an asymptotically zero-rate Steane code; the threshold of this code is found to be approximately $18.25\%$, in agreement with results from~\cite{harris_int_opt}. $1000$ Monte Carlo trials were utilized for the erasure and depolarizing noise simulations. Results are gathered per layer of concatenation $L$. The code snippets related to both (a) and (b) can be found in \cref{code1,code2}, respectively. Regarding computation time, for (a) the recovery rate curves for $L=1$ to $L=5$ took $14$\,s, $15$\,s, $18$\,s, $536$\,s, and $2204$\,s respectively. In (b) the cases $L=0$ to $L=2$ took $15$\,s, $194$\,s, and $1497$\,s, respectively (on an i9-14900K processor using 24 threads).}
\label{fig:decoder_results}
\end{figure*}

This section contains a step-by-step set of examples for utilizing \texttt{LEGO\_HQEC} as well as a runtime analysis. We begin by initializing and testing a maximum-rate \emph{Harlow-Preskill-Pastawski-Yoshida} (HaPPY)~\cite{preskill_happy_code} code under erasure noise, and then following up with an asymptotically zero-rate Steane code~\cite{harris_css_hqec} under Pauli depolarizing noise. The results of both decoder simulations are shown in \cref{fig:decoder_results}.

%--------------------------------------------%
\subsection{Usage}\label{section:example_usage}

%--------------------------------------------%
\subsubsection{Maximum-Rate HaPPY Code under Erasure Noise}\label{section:happy_erasure}

\begin{figure*}
\begin{lstlisting}[language=Python,caption={Code snippet for initializing the stabilizers and logical operators of the maximum-rate HaPPY code, as well as for testing the first through fifth layers of concatenation under the erasure noise channel. Decoding was carried out using the GE decoder.},  label=code1, captionpos=b]
from LEGO_HQEC.QuDec.ErasureDecoder import calculate_recovery_rates_for_p_range
from LEGO_HQEC.OperatorPush.Presets.HaPPY_code import setup_max_rate_happy
from LEGO_HQEC.OperatorPush.PushingToolbox import batch_push
from LEGO_HQEC.QuDec.InputProcessor import extract_stabilizers_from_result_dict, extract_logicals_from_result_dict
from LEGO_HQEC.QuDec.OutputProcessor import save_results_to_csv

if __name__ == '__main__':
    # Examine the HaPPY code with radii R=0, 1, 2, and 3, respectively.
    for R in [1, 2, 3, 4, 5]:
        # Obtain TN of the zero-rate HaPPY code with corresponding radius via preset.
        tensor_list = setup_max_rate_happy(R=R)
        results_dict = batch_push(tensor_list)
        stabilizers = extract_stabilizers_from_result_dict(results_dict)
        logical_zs, logical_xs = extract_logicals_from_result_dict(results_dict)
        logical_operators = [logical_zs[0]] + [logical_xs[0]]
        # Starting from an erasure error rate p of 0.0 to 1.0 with a step size of 0.05, 
        results = calculate_recovery_rates_for_p_range(n=1000, p_start=0.0, p_end=1.0, p_step=0.05, stabilizers=stabilizers, logical_operators=logical_operators)

        save_results_to_csv(results, file_path=F'R{R}_rec.csv')
\end{lstlisting}
\end{figure*}

\cref{code1} displays a code snippet for the construction and execution of the threshold calculation for the maximum-rate HaPPY code under erasure noise, utilizing the GE decoder discussed in \cref{section:decoders}. In the example above, the commensurate libraries and dependencies are first imported (lines 1-5), and then the stabilizers and logical operators for each layer are generated (lines 11-13). Those objects are then fed into the erasure decoder per layer $L$, and the Monte Carlo trials commence layerwise (line 17). Finally, the results are written into a CSV file and saved (line 19).  

%--------------------------------------------%
\subsubsection{Asymptotically Zero-Rate Steane Code under Depolarizing Noise}\label{section:steane_depo}

\begin{figure*}
\begin{lstlisting}[language=Python, caption={Code snippet for initializing the stabilizers and logical operators of the asymptotically zero-rate Steane code, as well as for testing the first three layers of concatenation under the Pauli depolarizing noise channel using the tensor-network decoder.}, label=code2, captionpos=b]
from LEGO_HQEC.QuDec.TN_decoder import tn_quantum_error_correction_decoder_multiprocess
from LEGO_HQEC.OperatorPush.Presets.Holographic_Steane_code import setup_heptagon_zero_rate_steane
from LEGO_HQEC.QuDec.OutputProcessor import save_results_to_csv
import numpy as np

if __name__ == '__main__':
    rx = 1/3
    rz = 1/3
    ry = 1/3
    for R in [0, 1, 2]:
        tensor_list = setup_heptagon_zero_rate_steane(R=R)
        p_depo_step = 0.05
        p_depo_start = 0.0
        p_depo_end = 0.5
        succ_rates = []
        for p_depo in np.arange(p_depo_start, p_depo_end + p_depo_step, p_depo_step):
            print(f"Running at R = {R}, p_depo = {p_depo}")
            succ_rate = tn_quantum_error_correction_decoder_multiprocess(tensor_list=tensor_list, p=p_depo, rx=rx, ry=ry, rz=rz, N=1000, n_process=8)
            succ_rates.append((p_depo, succ_rate))
        save_results_to_csv(results=succ_rates, file_path=f'R_{R}_depo.csv')

\end{lstlisting}
\end{figure*}

\cref{code2} shows the code needed for initializing the tensor-network decoder for a depolarizing noise channel. In contrast to the previous case, the tensor network decoder requires a specification of the relative proportions of $X$, $Y$, and $Z$ errors (lines 7-9). However, the actual generation of the stabilizers and logical operators proceeds identically as for the HaPPY code (lines 14-17). 

%--------------------------------------------%
\subsection{Runtime Performance}\label{section:runtime}

Here we report an example of the runtime performance expected of an asymptotically zero-rate HaPPY code for: the stabilizer and logical operator generation \texttt{OperatorPush}; the GE erasure decoder; the integer-optimization decoder; and the tensor-network decoder. All results are given as a function of layer $L$ and times are reported in seconds (s). For the decoders, the plotted results are \emph{per Monte Carlo trial}, while the \texttt{OperatorPush} results denote the per-layer start-to-finish time for pushing all locally-defined stabilizers and logical operators to their respective boundary representations, and extracting them into a CSV file. 

\begin{figure}
\centering
\includegraphics[width=\columnwidth]{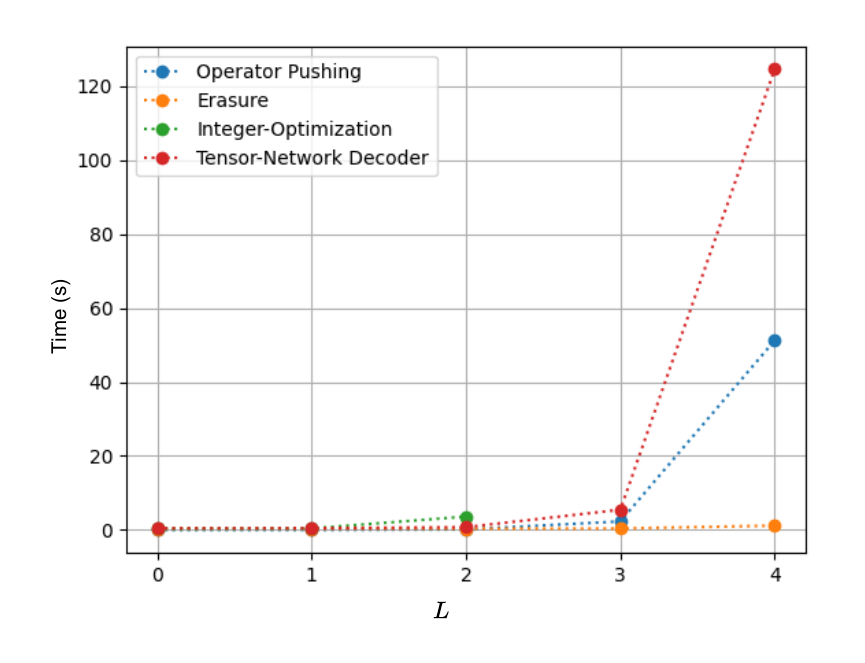}
\caption{Time-scaling results per event for the \texttt{OperatorPush} function, as well as the three different decoders utilized in this software package. Due to limitations in computational power, we show the results regarding the exponentially-scaling integer-optimization algorithm for only the first two layers of the asymptotically zero-rate HaPPY code.}
\label{fig:runtime_results}
\end{figure}

We were able to extract results for only the first two layers using the integer-optimization decoder, as we were limited by computational power. For all other decoders and the \texttt{OperatorPush} method, however, we were able to obtain time-scaling estimates for up to $L = 4$, i.e., on the order of roughly several thousand boundary (physical) qubits, depending on the Schl\"{a}fli symbols $\{p,q\}$ associated with the commensurate hyperbolic tiling.

%--------------------------------------------%
\section{Results for Holographic Black Hole Codes}\label{section:novel_results}

As a further demonstration of the range of our software tool's usability, we investigate the threshold behavior of the black hole HaPPY code under the erasure channel, which was introduced in~\cite{heterogeneous_holo_qrm}. 

\begin{figure}
\centering
\includegraphics[width=1.0\columnwidth]{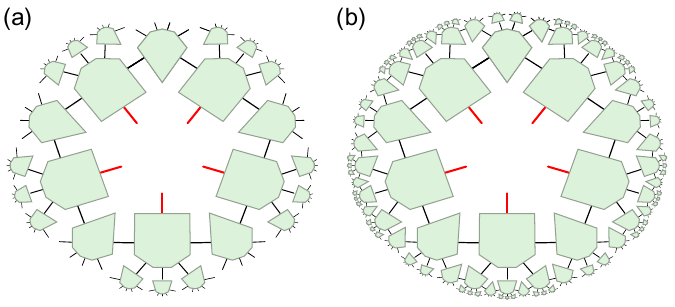}
\caption{The black hole HaPPY code with five logical qubits (red legs), constructed by removing the central tensor from the zero-rate code of \cref{fig:hqecc_types}(c). 
Here, we show the black hole code with (a) $L=2$ layers and (b) $L=3$ layers of edge inflation, corresponding to 95 and 375 physical qubits (outer boundary legs), respectively.
}
\label{fig:bh-code}
\end{figure}

\begin{figure}
\centering
\includegraphics[width=\columnwidth]{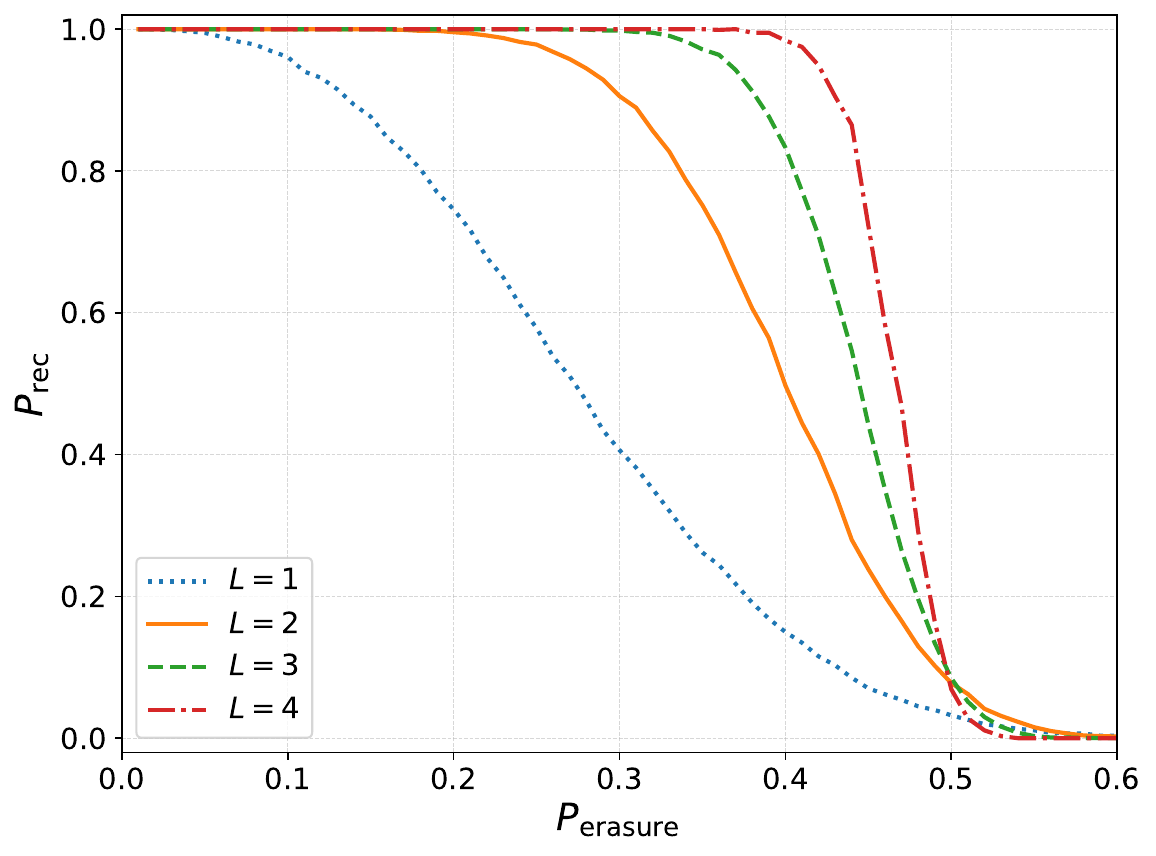}
\caption{Recovery probability of the black-hole HaPPY code under the erasure channel as a function of $P_{\text{erasure}}$ for $L\in \{1,\dots,4\}$. Each instance encodes five logical qubits, with success defined as simultaneous recovery of all. The threshold matches with that of the one-logical-qubit zero-rate HaPPY code, but with lower recovery probability. The black hole code is constructed by removing the tensor at layer $L=0$ from a regular holographic code (see Fig.~\ref{fig:bh-code}); hence, the smallest code starts from $L=1$. For the simulations, $L\in \{1,2,3\}$ used $10{,}000$ Monte Carlo trials for each $P_{\text{erasure}}$ value, and $L=4$ used $1{,}000$ trials, with step size $\Delta p_{\text{erasure}} = 0.01$. The computation times were $15$\,s, $141$\,s, $989$\,s, and $3805$\,s for $L\in \{1,\dots,4\}$, respectively (on an i9-14900K using 24 threads).}
\label{fig:bh_happy_erasure}
\end{figure}

Instances of the code are constructed by removing the central tensor and then growing $L$ edge inflation layers of the hexagon code tiling (see \cref{fig:bh-code}). Each instance encodes five logical qubits; stabilizers and logical operators are generated with \texttt{OperatorPush}, and decoding is performed using the Gaussian-elimination (GE) erasure decoder with $10^5$ Monte Carlo trials per $(p,L)$. Logical success probabilities are evaluated with respect to the simultaneous recoverability of all five logical qubits. We report a threshold $p_{\text{th}}=50\%$ from finite-size crossings (\cref{fig:bh_happy_erasure}). Consistent with our data, $p_{\text{th}}=50\%$ coincides with that of the zero-rate HaPPY code, while the black hole HaPPY code exhibits uniformly lower success probabilities; moreover, as $L$ increases the rate of the black hole HaPPY code approaches zero. Besides the black hole HaPPY code, we also constructed and tested the black hole holographic Steane code~\cite{harris2018calderbank} and the black hole holographic $[[6,1,3]]$ code~\cite{farrelly2021tensor}, obtaining similar threshold results around $p_{\text{th}} = 50\%$. These black hole codes are also asymptotically zero-rate; however, since they encode more logical qubits than their single-logical-qubit counterparts, their recovery curves lie slightly below while maintaining the same threshold.

%--------------------------------------------%
\section{Conclusion}\label{section:conclusion}

In this work, we have presented and demonstrated \texttt{LEGO\_HQEC}, a new software tool for constructing and testing holographic quantum codes under typical noise channels of interest. Similar to other notable software packages for quantum error correction, fault tolerance, among others~\cite{qgym_feld,sarkar2024yaqq,wille_qldpc,gidney2021stim,qecsim,QiskitCommunity2017}, our focus with \texttt{LEGO\_HQEC} has been to facilitate open-source development and enable future scientific discovery in the realm of holographic quantum error correction. In the process, we have: developed and implemented an easy-to-use open-source software tool; demonstrated several examples of its capabilities; and have provided a basic framework upon which more detailed code studies and further discoveries in holographic quantum error correction can be leveraged.

\texttt{LEGO\_HQEC} has already featured in several holographic code studies, including those exploring the novel \emph{Evenbly codes}~\cite{steinberg2024far,steinberg_2023}, analyses of asymptotically zero-rate holographic codes as compared to the \emph{hashing bound}~\cite{fan2024overcoming,junyu_thesis}, as well as for constructions of \emph{heterogeneous} holographic codes~\cite{heterogeneous_holo_qrm}, which are constructed from several seed codes and can circumvent the Eastin-Knill theorem regarding codes with universal logical gate sets~\cite{eastin2009restrictions,jochym2014using,chamberland2016thresholds}.

As a proof of concept, we presented the black-hole HaPPY code and its threshold under the erasure noise channel. Our numerical analysis shows that, while its threshold under the erasure channel coincides with that of the one-logical-qubit zero-rate HaPPY code, the overall recovery performance is reduced for finite code instances: Though the asymptotic rate is zero for both codes, comparing instances at a fixed number of physical qubits leads to higher rate and lower logical recovery probability for the black-hole code.  

There are many ideas for future modification of \texttt{LEGO\_HQEC}. Firstly, an extension for considering any hyperbolic tiling parameterized by $\{p,q\}$ would be enormously helpful, as it has been shown that certain classes of holographic codes have vastly different threshold behavior as the tiling is altered~\cite{steinberg2024far}. Furthermore, recent work~\cite{heterogeneous_holo_qrm,junyu_thesis} shows that incorporating multiple seed codes can allow for potentially \emph{universal fault-tolerant logical operations}~\cite{heterogeneous_holo_qrm}; tools to support such fault tolerance studies could be very useful. One can also consider implementing functionality for greater control over the encoded logical qubits in the bulk, as recent work shows that recovery rates of finite-rate holographic codes greatly depend not only on the number $k$ of encoded logical qubits, but critically on the \emph{positions} of said logical qubits in the bulk~\cite{alex_logical_recovery}. Even more broadly, it is known that the qLEGO formalism applies to \emph{non-Abelian} stabilizer codes, such as the \emph{XP stabilizer} variety~\cite{xp_stab_quantum_cao,webster2022xp}, or even \emph{non-stabilizer} codes, such as those presented in~\cite{exotic_gatesets,lidar2013quantum,non_additive1,non_additive2,non_additive3}; one may consider incorporating functionality to accommodate such holographic code constructions in the near future. Moreover, the class of codes known as \emph{Evenbly codes}, a derivate of the \emph{hyperinvariant tensor networks} first introduced in~\cite{hyper_evenbly,steinberg2022conformal}, have been extensively studied using \texttt{LEGO\_HQEC}~\cite{steinberg2024far,steinberg_2023}; however, one major impediment was the inclusion of \emph{Hadamard gates} on the contracted edges of the tensor network, thereby hindering decoding of Pauli channels with the tensor-network decoder. Allowing for such constraints, among others, remains within the future goals of this work. Finally, there are many potential additions purely from a software and usability standpoint, such as: adding functionality for \emph{generalized concatenated codes}~\cite{knill2005quantum,gottesman1997stabilizer,lidar2013quantum}; visualization tools, such as those that have recently appeared in the software package \texttt{PlanqTN}~\cite{pato_planqTN}; or GPU acceleration tools; among other prospects.  

%--------------------------------------------%
\subsection{Acknowledgements}
We thank David Elkouss and Rob Harris for useful discussions. MS and SF acknowledge financial support from the Intel corporation.  AJ is supported by the Einstein Research Unit ``Perspectives of a quantum digital transformation'' and Berlin Quantum. CC acknowledges support from the Commonwealth Cyber Initiative at Virginia Tech.

%--------------------------------------------%
\subsection{Author Contributions}
JF and MS implemented the Gaussian elimination decoder. JF developed the integer-optimization decoder, tensor-network decoder, and the \texttt{OperatorPush} generator during his master thesis; the master thesis itself was supervised by MS and SF. AS provided assistance with software design. MS wrote the manuscript, with the hyperbolic tiling figures provided by AJ. CC, AJ, AS, and SF provided project guidance and supervision.

%--------------------------------------------%
% \clearpage
\bibliographystyle{IEEEtran}
\bibliography{bibliography}

@article{steinberg_2023,
title={Holographic codes from hyperinvariant tensor networks},
volume={14},
ISSN={2041-1723},
url={http://dx.doi.org/10.1038/s41467-023-42743-z},
DOI={10.1038/s41467-023-42743-z},
number={7314},
journal={Nature Communications},
publisher={Springer Science and Business Media LLC},
author={Steinberg, Matthew and Feld, Sebastian and Jahn, Alexander},
year={2023},
month=nov }

@book{wilde2013quantum,
title={Quantum information theory},
author={Wilde, Mark M},
year={2013},
publisher={Cambridge university press}
}

@book{nielsen2010quantum,
title={Quantum computation and quantum information},
author={Nielsen, Michael A and Chuang, Isaac L},
year={2010},
publisher={Cambridge university press}
}

@article{capacities_erasure,
title = {Capacities of Quantum Erasure Channels},
author = {Bennett, Charles H. and DiVincenzo, David P. and Smolin, John A.},
journal = {Phys. Rev. Lett.},
volume = {78},
issue = {16},
pages = {3217--3220},
numpages = {0},
year = {1997},
month = {Apr},
publisher = {American Physical Society},
doi = {10.1103/PhysRevLett.78.3217},
url = {https://link.aps.org/doi/10.1103/PhysRevLett.78.3217}
}

@article{steinberg2024far,
title={Far from Perfect: Quantum Error Correction with (Hyperinvariant) Evenbly Codes},
author={Matthew Steinberg and Fan, Junyu and Harris, Robert J and Elkouss, David and Feld, Sebastian and Jahn, Alexander},
journal={arXiv preprint arXiv:2407.11926},
year={2024}
}

@article{harris_int_opt,
title = {Decoding holographic codes with an integer optimization decoder},
author = {Harris, Robert J. and Coupe, Elliot and McMahon, Nathan A. and Brennen, Gavin K. and Stace, Thomas M.},
journal = {Phys. Rev. A},
volume = {102},
issue = {6},
pages = {062417},
numpages = {6},
year = {2020},
month = {Dec},
publisher = {American Physical Society},
doi = {10.1103/PhysRevA.102.062417},
url = {https://link.aps.org/doi/10.1103/PhysRevA.102.062417}
}

@article{harris_css_hqec,
title = {Calderbank-Shor-Steane holographic quantum error-correcting codes},
author = {Harris, Robert J. and McMahon, Nathan A. and Brennen, Gavin K. and Stace, Thomas M.},
journal = {Phys. Rev. A},
volume = {98},
issue = {5},
pages = {052301},
numpages = {6},
year = {2018},
month = {Nov},
publisher = {American Physical Society},
doi = {10.1103/PhysRevA.98.052301},
url = {https://link.aps.org/doi/10.1103/PhysRevA.98.052301}
}

@article{farrelly_tn_codes,
title = {Tensor-Network Codes},
author = {Farrelly, Terry and Harris, Robert J. and McMahon, Nathan A. and Stace, Thomas M.},
journal = {Phys. Rev. Lett.},
volume = {127},
issue = {4},
pages = {040507},
numpages = {6},
year = {2021},
publisher = {American Physical Society},
doi = {10.1103/PhysRevLett.127.040507},
url = {https://link.aps.org/doi/10.1103/PhysRevLett.127.040507}
}

@article{Farrelly_ltnc,
title={Local tensor-network codes},
volume={24},
ISSN={1367-2630},
url={http://dx.doi.org/10.1088/1367-2630/ac5e87},
DOI={10.1088/1367-2630/ac5e87},
number={4},
journal={New Journal of Physics},
publisher={IOP Publishing},
author={Farrelly, Terry and Tuckett, David K and Stace, Thomas M},
year={2022},
month=apr, pages={043015} }

@article{preskill_happy_code,
title={Holographic quantum error-correcting codes: Toy models for the bulk/boundary correspondence},
author={Pastawski, Fernando and Yoshida, Beni and Harlow, Daniel and Preskill, John},
journal={Journal of High Energy Physics},
volume={2015},
number={6},
pages={1--55},
year={2015},
publisher={Springer}
}

@article{fan2024overcoming,
title={Overcoming the Zero-Rate Hashing Bound with Holographic Quantum Error Correction},
author={Junyu Fan and Matthew Steinberg and Jahn, Alexander and Cao, Charles and Feld, Sebastian},
journal={arXiv preprint arXiv:2408.06232},
year={2024}
}

@misc{heterogeneous_holo_qrm,
title={Universal fault-tolerant logic with heterogeneous holographic codes}, 
author={Matthew Steinberg and Junyu Fan and Jens Eisert and Sebastian Feld and Alexander Jahn and Chunjun Cao},
year={2025},
eprint={2504.10386},
archivePrefix={arXiv},
primaryClass={quant-ph},
url={https://arxiv.org/abs/2504.10386}, 
}

@article{alex_logical_recovery,
author = {Alexander Jahn and Matthew Steinberg and Jens Eisert},
title = {Holographic Codes with Many Logical Qubits},
journal = {In Preparation},
year = {2025}
}

@article{xp_stab_quantum_cao,
title={Quantum Lego and XP Stabilizer Codes},
author={Shen, Ruohan and Wang, Yixu and Cao, ChunJun},
journal={arXiv preprint arXiv:2310.19538},
year={2023}
}

@inproceedings{penrose1955generalized,
title={A generalized inverse for matrices},
author={Penrose, Roger},
booktitle={Mathematical proceedings of the Cambridge philosophical society},
volume={51},
pages={406--413},
year={1955},
organization={Cambridge University Press}
}

@article{moore1920reciprocal,
title={On the reciprocal of the general algebraic matrix},
author={Moore, Eliakim H},
journal={Bulletin of the american mathematical society},
volume={26},
pages={294--295},
year={1920}
}

@article{gottesman_improved_sim,
title = {Improved simulation of stabilizer circuits},
author = {Aaronson, Scott and Gottesman, Daniel},
journal = {Phys. Rev. A},
volume = {70},
issue = {5},
pages = {052328},
numpages = {14},
year = {2004},
month = {Nov},
publisher = {American Physical Society},
doi = {10.1103/PhysRevA.70.052328},
url = {https://link.aps.org/doi/10.1103/PhysRevA.70.052328}
}

@article{farrelly_parallel_decoding,
title = {Parallel decoding of multiple logical qubits in tensor-network codes},
author = {Farrelly, Terry and Milicevic, Nicholas and Harris, Robert J. and McMahon, Nathan A. and Stace, Thomas M.},
journal = {Phys. Rev. A},
volume = {105},
issue = {5},
pages = {052446},
numpages = {12},
year = {2022},
month = {May},
publisher = {American Physical Society},
doi = {10.1103/PhysRevA.105.052446},
url = {https://link.aps.org/doi/10.1103/PhysRevA.105.052446}
}

@article{steanecode1,
title = {Good quantum error-correcting codes exist},
author = {Calderbank, A. R. and Shor, Peter W.},
journal = {Phys. Rev. A},
volume = {54},
issue = {2},
pages = {1098--1105},
numpages = {0},
year = {1996},
publisher = {American Physical Society},
doi = {10.1103/PhysRevA.54.1098},
url = {https://link.aps.org/doi/10.1103/PhysRevA.54.1098}
}

@article{steanecode2,
title = {Simple quantum error-correcting codes},
author = {Steane, A. M.},
journal = {Phys. Rev. A},
volume = {54},
issue = {6},
pages = {4741--4751},
numpages = {0},
year = {1996},
publisher = {American Physical Society},
doi = {10.1103/PhysRevA.54.4741},
url = {https://link.aps.org/doi/10.1103/PhysRevA.54.4741}
}

@article{laflamme1996perfect,
title={Perfect quantum error correcting code},
author={Laflamme, Raymond and Miquel, Cesar and Paz, Juan Pablo and Zurek, Wojciech Hubert},
journal={Physical Review Letters},
volume={77},
number={1},
pages={198},
year={1996},
publisher={APS}
}

@article{quantumlego,
title = {Quantum Lego: Building Quantum Error Correction Codes from Tensor Networks},
author = {Cao, ChunJun and Lackey, Brad},
journal = {PRX Quantum},
volume = {3},
issue = {2},
pages = {020332},
numpages = {37},
year = {2022},
publisher = {American Physical Society},
doi = {10.1103/PRXQuantum.3.020332},
url = {https://link.aps.org/doi/10.1103/PRXQuantum.3.020332}
}

@mastersthesis{junyu_thesis,
author = {Junyu Fan},
title = {Biased-Noise Threshold Studies for Holographic Quantum Error-Correction Codes},
school = {QuTech, Technical University of Delft},
address = {Delft, the Netherlands},
year = {2024}
}

@article{chubb2021statistical,
title={Statistical mechanical models for quantum codes with correlated noise},
author={Chubb, Christopher T and Flammia, Steven T},
journal={Annales de l’Institut Henri Poincar{\'e} D},
volume={8},
number={2},
pages={269--321},
year={2021}
}

@inproceedings{qgym_feld,
title={qgym: A Gym for training and benchmarking RL-based quantum compilation},
author={Van Der Linde, Stan and De Kok, Willem and Bontekoe, Tariq and Feld, Sebastian},
booktitle={2023 IEEE International Conference on Quantum Computing and Engineering (QCE)},
volume={2},
pages={26--30},
year={2023},
organization={IEEE}
}

@inproceedings{wille_qldpc,
title={Software tools for decoding quantum low-density parity-check codes},
author={Berent, Lucas and Burgholzer, Lukas and Wille, Robert},
booktitle={Proceedings of the 28th Asia and South Pacific Design Automation Conference},
pages={709--714},
year={2023}
}

@article{steinberg2022conformal,
title={Conformal properties of hyperinvariant tensor networks},
author={Steinberg, Matthew and Prior, Javier},
journal={Scientific Reports},
volume={12},
number={1},
pages={532},
year={2022},
publisher={Nature Publishing Group UK London}
}

@article{hyper_evenbly,
title = {Hyperinvariant Tensor Networks and Holography},
author = {Evenbly, Glen},
journal = {Phys. Rev. Lett.},
volume = {119},
issue = {14},
pages = {141602},
numpages = {5},
year = {2017},
month = {Oct},
publisher = {American Physical Society},
doi = {10.1103/PhysRevLett.119.141602},
url = {https://link.aps.org/doi/10.1103/PhysRevLett.119.141602}
}

@article{gidney2021stim,
title={Stim: a fast stabilizer circuit simulator},
author={Gidney, Craig},
journal={Quantum},
volume={5},
pages={497},
year={2021},
publisher={Verein zur F{\"o}rderung des Open Access Publizierens in den Quantenwissenschaften}
}

@misc{QiskitCommunity2017,
title     = {Qiskit: {{An}} Open-Source Framework for Quantum Computing},
author    = {{Qiskit Community}},
year      = {2017},
month     = mar,
doi       = {10.5281/zenodo.2562110},
url       = {https://github.com/Qiskit/qiskit}
}

@phdthesis{qecsim,
author = {Tuckett, David Kingsley},
title = {Tailoring surface codes: Improvements in quantum error correction with biased noise},
school = {University of Sydney},
doi = {10.25910/x8xw-9077},
year = {2020},
note = {(qecsim: \url{https://github.com/qecsim/qecsim})}
}

@article{exotic_gatesets,
title = {Family of Quantum Codes with Exotic Transversal Gates},
author = {Kubischta, Eric and Teixeira, Ian},
journal = {Phys. Rev. Lett.},
volume = {131},
issue = {24},
pages = {240601},
numpages = {6},
year = {2023},
month = {Dec},
publisher = {American Physical Society},
doi = {10.1103/PhysRevLett.131.240601},
url = {https://link.aps.org/doi/10.1103/PhysRevLett.131.240601}
}

@article{conformal_quasicrystals_holography,
title = {Conformal Quasicrystals and Holography},
author = {Boyle, Latham and Dickens, Madeline and Flicker, Felix},
journal = {Phys. Rev. X},
volume = {10},
issue = {1},
pages = {011009},
numpages = {10},
year = {2020},
month = {Jan},
publisher = {American Physical Society},
doi = {10.1103/PhysRevX.10.011009},
url = {https://link.aps.org/doi/10.1103/PhysRevX.10.011009}
}

@article{central_charge_aperiodic,
title = {Central charges of aperiodic holographic tensor-network models},
author = {Jahn, Alexander and Zimbor\'as, Zolt\'an and Eisert, Jens},
journal = {Phys. Rev. A},
volume = {102},
issue = {4},
pages = {042407},
numpages = {18},
year = {2020},
month = {Oct},
publisher = {American Physical Society},
doi = {10.1103/PhysRevA.102.042407},
url = {https://link.aps.org/doi/10.1103/PhysRevA.102.042407}
}

@article{jahn2022tensor_qcft,
title={Tensor network models of AdS/qCFT},
author={Jahn, Alexander and Zimbor{\'a}s, Zolt{\'a}n and Eisert, Jens},
journal={Quantum},
volume={6},
pages={643},
year={2022},
publisher={Verein zur F{\"o}rderung des Open Access Publizierens in den Quantenwissenschaften}
}

@article{jahn_topical,
title={Holographic tensor network models and quantum error correction: a topical review},
author={Jahn, Alexander and Eisert, Jens},
journal={Quantum Science and Technology},
volume={6},
number={3},
pages={033002},
year={2021},
publisher={IOP Publishing}
}

@article{joe_chubb,
title = {Tensor-Network Decoding Beyond 2D},
author = {Piveteau, Christophe and Chubb, Christopher T. and Renes, Joseph M.},
journal = {PRX Quantum},
volume = {5},
issue = {4},
pages = {040303},
numpages = {20},
year = {2024},
month = {Oct},
publisher = {American Physical Society},
doi = {10.1103/PRXQuantum.5.040303},
url = {https://link.aps.org/doi/10.1103/PRXQuantum.5.040303}
}

@article{campbell2017roadsterhal,
title={Roads towards fault-tolerant universal quantum computation},
author={Campbell, Earl T and Terhal, Barbara M and Vuillot, Christophe},
journal={Nature},
volume={549},
number={7671},
pages={172--179},
year={2017},
publisher={Nature Publishing Group UK London}
}

@article{terhal2015quantum,
title={Quantum error correction for quantum memories},
author={Terhal, Barbara M},
journal={Reviews of Modern Physics},
volume={87},
number={2},
pages={307--346},
year={2015},
publisher={APS}
}

@article{yoshida2024concatenate,
title={Concatenate codes, save qubits},
author={Yoshida, Satoshi and Tamiya, Shiro and Yamasaki, Hayata},
journal={arXiv preprint arXiv:2402.09606},
year={2024}
}

@article{harlow2015bulk,
title={Bulk locality and quantum error correction in AdS/CFT},
author={Almheiri, Ahmed and Dong, Xi and Harlow, Daniel},
journal={Journal of High Energy Physics},
volume={2015},
number={4},
pages={1--34},
year={2015},
publisher={Springer}
}

@article{maldacena1999large,
title={The large-N limit of superconformal field theories and supergravity},
author={Maldacena, Juan},
journal={International journal of theoretical physics},
volume={38},
number={4},
pages={1113--1133},
year={1999},
publisher={Springer}
}

@article{witten1998anti,
title={Anti de Sitter space and holography},
author={Witten, Edward},
journal={arXiv preprint hep-th/9802150},
year={1998}
}

@article{gubser1998gauge,
title={Gauge theory correlators from non-critical string theory},
author={Gubser, Steven S and Klebanov, Igor R and Polyakov, Alexander M},
journal={Physics Letters B},
volume={428},
number={1-2},
pages={105--114},
year={1998},
publisher={Elsevier}
}

@article{ratcliffe1994foundations,
title={Foundations of Hyperbolic Manifolds},
author={Ratcliffe, JG},
journal={Graduate Texts in Mathematics/Springer-Verlag},
volume={149},
year={1994}
}

@article{fowler2012surface,
title={Surface codes: Towards practical large-scale quantum computation},
author={Fowler, Austin G and Mariantoni, Matteo and Martinis, John M and Cleland, Andrew N},
journal={Physical Review A—Atomic, Molecular, and Optical Physics},
volume={86},
number={3},
pages={032324},
year={2012},
publisher={APS}
}

@article{gottesman2013fault,
title={Fault-tolerant quantum computation with constant overhead},
author={Gottesman, Daniel},
journal={arXiv preprint arXiv:1310.2984},
year={2013}
}

@article{gottesmanFT,
title = {Theory of fault-tolerant quantum computation},
author = {Gottesman, Daniel},
journal = {Phys. Rev. A},
volume = {57},
issue = {1},
pages = {127--137},
numpages = {0},
year = {1998},
month = {Jan},
publisher = {American Physical Society},
doi = {10.1103/PhysRevA.57.127},
url = {https://link.aps.org/doi/10.1103/PhysRevA.57.127}
}

@article{preskill2018quantum,
title={Quantum computing in the NISQ era and beyond},
author={Preskill, John},
journal={Quantum},
volume={2},
pages={79},
year={2018},
publisher={Verein zur F{\"o}rderung des Open Access Publizierens in den Quantenwissenschaften}
}

@book{lidar2013quantum,
title={Quantum error correction},
author={Lidar, Daniel A and Brun, Todd A},
year={2013},
publisher={Cambridge university press}
}

@book{gottesman1997stabilizer,
title={Stabilizer codes and quantum error correction},
author={Gottesman, Daniel},
year={1997},
publisher={California Institute of Technology}
}

@article{breuckmann2021quantum,
title={Quantum low-density parity-check codes},
author={Breuckmann, Nikolas P and Eberhardt, Jens Niklas},
journal={PRX Quantum},
volume={2},
number={4},
pages={040101},
year={2021},
publisher={APS}
}

@article{albert2022bosonic,
title={Bosonic coding: introduction and use cases},
author={Albert, Victor V},
journal={arXiv preprint arXiv:2211.05714},
year={2022}
}

@article{between_shorsteane,
title = {Between Shor and Steane: A Unifying Construction for Measuring Error Syndromes},
author = {Huang, Shilin and Brown, Kenneth R.},
journal = {Phys. Rev. Lett.},
volume = {127},
issue = {9},
pages = {090505},
numpages = {5},
year = {2021},
month = {Aug},
publisher = {American Physical Society},
doi = {10.1103/PhysRevLett.127.090505},
url = {https://link.aps.org/doi/10.1103/PhysRevLett.127.090505}
}

@inproceedings{dhruvi_groovy,
title={Low-depth flag-style syndrome extraction for small quantum error-correction codes},
author={Bhatnagar, Dhruv and Steinberg, Matthew and Elkouss, David and Almudever, Carmen G and Feld, Sebastian},
booktitle={2023 IEEE International Conference on Quantum Computing and Engineering (QCE)},
volume={1},
pages={63--69},
year={2023},
organization={IEEE}
}

@article{chamberland2018flag,
title={Flag fault-tolerant error correction with arbitrary distance codes},
author={Chamberland, Christopher and Beverland, Michael E},
journal={Quantum},
volume={2},
pages={53},
year={2018},
publisher={Verein zur F{\"o}rderung des Open Access Publizierens in den Quantenwissenschaften}
}

@article{delfosse2021beyond,
title={Beyond single-shot fault-tolerant quantum error correction},
author={Delfosse, Nicolas and Reichardt, Ben W and Svore, Krysta M},
journal={IEEE Transactions on Information Theory},
volume={68},
number={1},
pages={287--301},
year={2021},
publisher={IEEE}
}

@article{bombin2015single,
title={Single-shot fault-tolerant quantum error correction},
author={Bomb{\'\i}n, H{\'e}ctor},
journal={Physical Review X},
volume={5},
number={3},
pages={031043},
year={2015},
publisher={APS}
}

@article{bombin2015gauge,
title={Gauge color codes: optimal transversal gates and gauge fixing in topological stabilizer codes},
author={Bomb{\'\i}n, H{\'e}ctor},
journal={New Journal of Physics},
volume={17},
number={8},
pages={083002},
year={2015},
publisher={IOP Publishing}
}

@article{divincenzo2007effective,
title={Effective fault-tolerant quantum computation with slow measurements},
author={DiVincenzo, David P and Aliferis, Panos},
journal={Physical review letters},
volume={98},
number={2},
pages={020501},
year={2007},
publisher={APS}
}

@article{divincenzo1996fault,
title={Fault-tolerant error correction with efficient quantum codes},
author={DiVincenzo, David P and Shor, Peter W},
journal={Physical review letters},
volume={77},
number={15},
pages={3260},
year={1996},
publisher={APS}
}

@article{chao2018quantum,
title={Quantum error correction with only two extra qubits},
author={Chao, Rui and Reichardt, Ben W},
journal={Physical review letters},
volume={121},
number={5},
pages={050502},
year={2018},
publisher={APS}
}

@article{knill2005quantum,
title={Quantum computing with realistically noisy devices},
author={Knill, Emanuel},
journal={Nature},
volume={434},
number={7029},
pages={39--44},
year={2005},
publisher={Nature Publishing Group UK London}
}

@article{aliferis2005quantum,
title={Quantum accuracy threshold for concatenated distance-3 codes},
author={Aliferis, Panos and Gottesman, Daniel and Preskill, John},
journal={arXiv preprint quant-ph/0504218},
year={2005}
}

@inproceedings{shor1996fault,
title={Fault-tolerant quantum computation},
author={Shor, Peter W},
booktitle={Proceedings of 37th conference on foundations of computer science},
pages={56--65},
year={1996},
organization={IEEE}
}

@article{delfosse2020short,
title={Short shor-style syndrome sequences},
author={Delfosse, Nicolas and Reichardt, Ben W},
journal={arXiv preprint arXiv:2008.05051},
year={2020}
}

@article{iolius2023decoding,
title={Decoding algorithms for surface codes},
author={iOlius, Antonio deMarti and Fuentes, Patricio and Or{\'u}s, Rom{\'a}n and Crespo, Pedro M and Martinez, Josu Etxezarreta},
journal={arXiv preprint arXiv:2307.14989},
year={2023}
}

@inproceedings{dinur2023good,
title={Good quantum LDPC codes with linear time decoders},
author={Dinur, Irit and Hsieh, Min-Hsiu and Lin, Ting-Chun and Vidick, Thomas},
booktitle={Proceedings of the 55th annual ACM symposium on theory of computing},
pages={905--918},
year={2023}
}

@article{roffe2020decoding,
title={Decoding across the quantum low-density parity-check code landscape},
author={Roffe, Joschka and White, David R and Burton, Simon and Campbell, Earl},
journal={Physical Review Research},
volume={2},
number={4},
pages={043423},
year={2020},
publisher={APS}
}

@article{babar2015fifteen,
title={Fifteen years of quantum LDPC coding and improved decoding strategies},
author={Babar, Zunaira and Botsinis, Panagiotis and Alanis, Dimitrios and Ng, Soon Xin and Hanzo, Lajos},
journal={iEEE Access},
volume={3},
pages={2492--2519},
year={2015},
publisher={IEEE}
}

@article{eastin2009restrictions,
title={Restrictions on transversal encoded quantum gate sets},
author={Eastin, Bryan and Knill, Emanuel},
journal={Physical review letters},
volume={102},
number={11},
pages={110502},
year={2009},
publisher={APS}
}

@article{jochym2014using,
title={Using concatenated quantum codes for universal fault-tolerant quantum gates},
author={Jochym-O’Connor, Tomas and Laflamme, Raymond},
journal={Physical review letters},
volume={112},
number={1},
pages={010505},
year={2014},
publisher={APS}
}

@article{chamberland2016thresholds,
title={Thresholds for universal concatenated quantum codes},
author={Chamberland, Christopher and Jochym-O’Connor, Tomas and Laflamme, Raymond},
journal={Physical review letters},
volume={117},
number={1},
pages={010501},
year={2016},
publisher={APS}
}

@inproceedings{zyczkowski_mme,
title={Maximally entangled multipartite states: a brief survey},
author={Enr{\'\i}quez, M and Wintrowicz, I and {\.Z}yczkowski, Karol},
booktitle={Journal of Physics: Conference Series},
volume={698},
pages={012003},
year={2016},
organization={IOP Publishing}
}

@article{bravyi2005universal,
title={Universal quantum computation with ideal Clifford gates and noisy ancillas},
author={Bravyi, Sergey and Kitaev, Alexei},
journal={Physical Review A—Atomic, Molecular, and Optical Physics},
volume={71},
number={2},
pages={022316},
year={2005},
publisher={APS}
}

@article{codeswitching1,
title = {Fault-Tolerant Code-Switching Protocols for Near-Term Quantum Processors},
author = {Butt, Friederike and Heu\ss{}en, Sascha and Rispler, Manuel and M\"uller, Markus},
journal = {PRX Quantum},
volume = {5},
issue = {2},
pages = {020345},
numpages = {26},
year = {2024},
month = {May},
publisher = {American Physical Society},
doi = {10.1103/PRXQuantum.5.020345},
url = {https://link.aps.org/doi/10.1103/PRXQuantum.5.020345}
}

@article{codeswitching2,
title = {Fault-Tolerant Conversion between the Steane and Reed-Muller Quantum Codes},
author = {Anderson, Jonas T. and Duclos-Cianci, Guillaume and Poulin, David},
journal = {Phys. Rev. Lett.},
volume = {113},
issue = {8},
pages = {080501},
numpages = {5},
year = {2014},
month = {Aug},
publisher = {American Physical Society},
doi = {10.1103/PhysRevLett.113.080501},
url = {https://link.aps.org/doi/10.1103/PhysRevLett.113.080501}
}

@article{codeswitching3,
title = {Universal transversal gates with color codes: A simplified approach},
author = {Kubica, Aleksander and Beverland, Michael E.},
journal = {Phys. Rev. A},
volume = {91},
issue = {3},
pages = {032330},
numpages = {12},
year = {2015},
month = {Mar},
publisher = {American Physical Society},
doi = {10.1103/PhysRevA.91.032330},
url = {https://link.aps.org/doi/10.1103/PhysRevA.91.032330}
}

@article{sarkar2024yaqq,
title={YAQQ: Yet Another Quantum Quantizer--Design Space Exploration of Quantum Gate Sets using Novelty Search},
author={Sarkar, Aritra and Kundu, Akash and Steinberg, Matthew and Mishra, Sibasish and Fauquenot, Sebastiaan and Acharya, Tamal and Miszczak, Jaros{\l}aw A and Feld, Sebastian},
journal={arXiv preprint arXiv:2406.17610},
year={2024}
}

@article{webster2022xp,
title={The XP stabiliser formalism: a generalisation of the Pauli stabiliser formalism with arbitrary phases},
author={Webster, Mark A and Brown, Benjamin J and Bartlett, Stephen D},
journal={Quantum},
volume={6},
pages={815},
year={2022},
publisher={Verein zur F{\"o}rderung des Open Access Publizierens in den Quantenwissenschaften}
}

@misc{pato_planqTN,
author = {Pato, Balint and Vanlerberghe, June and Cao, ChunJun and Lackey, Brad and Brown, Kenneth},
title = {PlanqTN, a Python library and interactive web app implementing the quantum LEGO framework},
month = aug,
year = 2025,
publisher = {Zenodo},
doi = {10.5281/zenodo.16761072},
url = {https://doi.org/10.5281/zenodo.16761072}
}

@article{non_additive1,
title={On the structure of additive quantum codes and the existence of nonadditive codes},
author={Roychowdhury, Vwani P and Vatan, Farrokh},
journal={arXiv preprint quant-ph/9710031},
year={1997}
}

@article{non_additive2,
title={A note on non-additive quantum codes},
author={Grassl, Markus and Beth, Thomas},
journal={arXiv preprint quant-ph/9703016},
year={1997}
}

@article{non_additive3,
title={A nonadditive quantum code},
author={Rains, Eric M and Hardin, RH and Shor, Peter W and Sloane, NJA},
journal={Physical Review Letters},
volume={79},
number={5},
pages={953},
year={1997},
publisher={APS}
}

@article{bravyi_suchara,
title = {Efficient algorithms for maximum likelihood decoding in the surface code},
author = {Bravyi, Sergey and Suchara, Martin and Vargo, Alexander},
journal = {Phys. Rev. A},
volume = {90},
issue = {3},
pages = {032326},
numpages = {15},
year = {2014},
month = {Sep},
publisher = {American Physical Society},
doi = {10.1103/PhysRevA.90.032326},
url = {https://link.aps.org/doi/10.1103/PhysRevA.90.032326}
}

@inproceedings{le_gall_matrix_mult,
title={Powers of tensors and fast matrix multiplication},
author={Le Gall, Fran{\c{c}}ois},
booktitle={Proceedings of the 39th international symposium on symbolic and algebraic computation},
pages={296--303},
year={2014}
}

@article{farrelly2021tensor,
  title={Tensor-network codes},
  author={Farrelly, Terry and Harris, Robert J and McMahon, Nathan A and Stace, Thomas M},
  journal={Physical Review Letters},
  volume={127},
  number={4},
  pages={040507},
  year={2021},
  publisher={APS}
}

@article{harris2018calderbank,
  title={Calderbank-Shor-Steane holographic quantum error-correcting codes},
  author={Harris, Robert J and McMahon, Nathan A and Brennen, Gavin K and Stace, Thomas M},
  journal={Physical Review A},
  volume={98},
  number={5},
  pages={052301},
  year={2018},
  publisher={APS}
}

\end{document}